\documentclass[runningheads,a4paper]{llncs}

\usepackage{amssymb}
\usepackage{color}
\usepackage{graphicx}

\newcommand{\keywords}[1]{\par\addvspace\baselineskip
\noindent\keywordname\enspace\ignorespaces#1}

\begin{document}

\mainmatter  

\title{A stochastic model of\\social interaction in wild house mice}

\titlerunning{A stochastic model of social interaction in wild house mice}

\author{Nicolas Perony$^1$\and Barbara K\"onig$^2$\and Frank Schweitzer$^1$}
\authorrunning{Nicolas Perony et al.}

\institute{
Chair of Systems Design, ETH Zurich,\\Weinbergstr. 56-58, CH-8092 Zurich, Switzerland
\and
Institute of Evolutionary Biology and Environmental Studies, University of
Zurich,\\Winterthurerstr. 190, CH-8057 Zurich, Switzerland\\
\url{nperony@ethz.ch}, \url{http://www.sg.ethz.ch}}

\maketitle

\begin{center}
\bf 
Proceedings of the European Conference on Complex Systems 2010\\
(ECCS'10), Lisbon, September 13-17, 2010
\end{center}

\begin{abstract}
We investigate to what extent the interaction dynamics of a population of wild
house mouse (\emph{Mus musculus domesticus}) in their environment can be explained
by a simple stochastic model. We use a Markov chain model to describe the transitions
of mice in a discrete space of nestboxes, and implement a multi-agent simulation of the
model. We find that some important features of our behavioural dataset can be
reproduced using this simplified stochastic representation, and discuss
the improvements that could be made to our model in order to increase the accuracy
of its predictions. Our findings have implications for the understanding of the
complexity underlying social behaviour in the animal kingdom and the cognitive
requirements of such behaviour.
\end{abstract}

\section{Introduction}

There is current debate as to what aspects of the social behaviour observed in an animal
species can be explained as a specific adaptation versus a side-effect of the interaction
of individuals with their environment. If some emerging properties of a group's social
structure result from simple behavioural mechanisms, it may be that what is often thought
to be an explicitly social behaviour does not require as much cognitive capacities as it
is assumed.

A great many animal species present a high level of cognitive development and a longly
evolved complex social structure (\cite{dewaal2003}). Although it has been long known that
living in groups has an adaptive value (\cite{krause2002}), little is known of the process
by which social bonds form between individuals, and how tightly-knit communities emerge
and eventually dissolve from those social bonds. In wild house mice ({\it Mus musculus
domesticus}), cooperative breeding, short generation time, polygynandry, high reproductive
and life expectancy skew in both sexes (\cite{lindholm2009,manser2009}), lead to
high flexibility in behaviour and social organisation. Therefore, this is a choice species
to study the dynamics of group formation and evolution. Here we address the question of what
aspects of the mice's social structure can be explained by the self-emergent properties of
collective random behaviour, and especially which patterns may rather result from social
adaptation produced by individual selection.

\section{Our study system}
We study an established free-living population of wild house mice in a barn outside of Zurich,
where mice can freely emigrate and immigrate. In the barn, the mice nest in 40 artificial
nestboxes, and are provided with straw as nesting material, and food and water outside the
boxes. At four to six week intervals, comprehensive trapping is conducted to monitor the
population, and adult mice are implanted with a transponder (RFID tag) so that they are individually
identifiable. Transponder readers are installed in the tunnels that provide entrances to the nestboxes;
these readers connect to a computer and continuously track movements into and out of nestboxes.
This provides us with 24-hour information on movements and social affiliations of adult mice.
Data collection started in May 2007 and totals around 14 million individual recordings as of February
2010. We study the period ranging from Jan. 1, 2008 to Dec. 31, 2009.

\begin{figure}
\centering
\includegraphics[width=5cm, angle=-90]{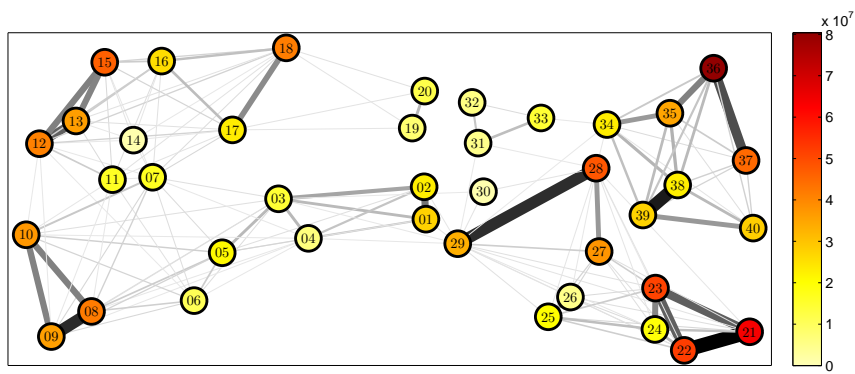}
\caption{Box occupation pattern and traffic between boxes: the colour of the boxes represents their
cumulated stay duration in seconds, while the darkness and thickness of the interbox edges represent
the intensity of the bidirectional traffic between the boxes (for clarity purposes, all edges with a
traffic $<50$ trips were omitted).}
\label{fig:boxoccupation}
\end{figure}

Our 2-year long dataset covers the 11'259'557 location records of 508 mice,
accounting for 1'376'720 stays in all 40 nestboxes, and leading to 1'064'695 one-to-one
encounters, whose frequency, context and duration we use as a proxy for the characterisation
of social interactions. Figure~\ref{fig:boxoccupation} shows the geographical positioning of the
nextboxes, as well as the heterogeneity of their occupation pattern: indeed, the total
occupation duration per box ranges from 264 to 22332 hours (the lowest figure can however
be attributed to a malfunctioning RFID antenna; the maximum value is longer than our study
period because some stays can overlap when several mice are together in a box). Hence, it is
possible to identify the ``hotspots'' of the barn and the busiest routes between nestboxes.

\section{A proposed model of social interaction}
Computer-based research in social science proceeds by building simplified representations of social
phenomena, often much too complex in the number of factors involved and the non-linearity of the dependencies
between those factors to allow for a thorough and exact description (\cite{gilbert2000}). We make the assumption
that through the collective behaviour of agents whose complexity lies far below that of real mice, we can
reproduce some of the global behavioural patterns we observe in the barn. We undertake a step-wise approach
towards a better understanding of the mechanisms governing the mice's social behaviour in the barn. To begin
with, we describe the system as a Markov chain and test the validity of the approach. We then further the
analysis by implementing our model in a multi-agent simulation calibrated from the real data. We use the
discrepancies of the model's output compared to the the observed patterns to point out the system's features
which may not be due to complex social behaviour. We discuss the accuracy of such a simple model in the
reproduction of behavioural patterns as well as its parametrisation using hypotheses from behavioural science.

\subsection{Markov chain model}
In this first step, we focus on a coarse-grained description of the problem: rather than following the individual
trajectories of all mice, we concentrate on the occupation of each nestbox. This has the advantage of reducing
the complexity of the approach by not having to describe all individual movement equations. Moreover, the number
of simulation variables thus stays constant (it depends solely on the number of nestboxes), whatever the number
of agents in the simulation is. In other terms, we simulate the behaviour of the agents (mice) but observe only
the result of their collective behaviour on the occupation density per box in the barn.

\begin{figure}
\centering
\includegraphics[width=11cm]{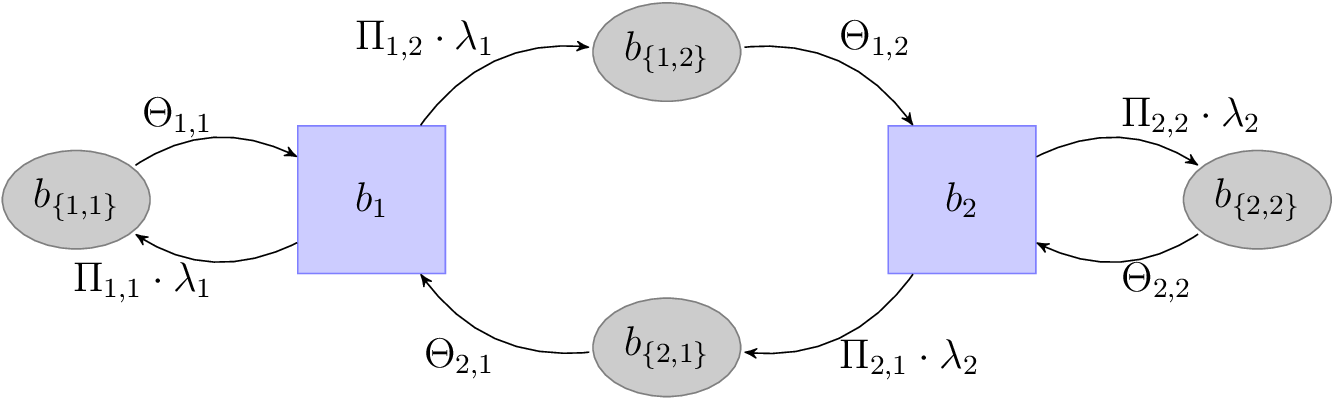}
\caption{Diagram of the states an agent can occupy in the case where the number of nestboxes $B=2$. $b_1$ and $b_2$ are the
2 nestboxes, to which are associated $B^2=4$ transit boxes corresponding to the intermediary states between any box and any
other box (including itself). Edges are labeled with the transition rates from state to state.}
\label{fig:markovchain1}
\end{figure}

We describe the transitions from state to state (or here, from nestbox to nestbox) by a Markov chain (see Fig.
~\ref{fig:markovchain1}). The transition rate of a Markov process in the chain is the probability for an agent
of leaving its nestbox multiplied by the transit frequency from this nestbox to any nestbox, i.e. the joint probability
of a mouse leaving its nestbox for any nestbox, including a return to the same nestbox. The Markov property of the
processes can be expressed at the agent level as follows: the conditional probability of an agent's future state depends
only on its current state, or in other words the mice keep no memory of the boxes they visited prior to the one they
currently stay in.

When going from a box to another, an agent needs a finite time. Moreover, we observe that for any 3 nestboxes $i,j,k$,
the transit times from $i$ to $j$ and from $i$ to $k$ are different ($\chi ^2$ test, normality hypothesis on the
distribution of transit times rejected at a 0.05 confidence level for 23 boxes out of 40). To cover this specificity
of our system, we need to introduce the concept of \textit{transit box} (see Fig.~\ref{fig:markovchain1}): whenever an
agent is not in a nestbox, we define that it is in an intermediary (transit) state between its origin and destination
nestboxes, whose leaving rate depends the particular origin and destination boxes. We regroup hereinafter the terms
``nestbox'' and ``transit box'' under the common denomination ``box''.

\subsubsection{Model description} The $N$ agents (the mice) are assumed to move within the discrete state-space
formed by the $B$ nestboxes and $B^2$ transit nestboxes according to stochastic processes. Therefore,
each agent has only one parameter, its current box $b$. All processes are assumed to have the Markov property. We can
thereby describe the agents' behaviour by a Markov chain, as illustrated in Fig.~\ref{fig:markovchain1}. $\lambda$,
$\Theta$ and $\Pi$ are fixed parameters of the system.

\begin{itemize}
 \item $\lambda$ is the \textit{leaving rate}, a vector of length $B$. This is the rate at which an agent leaves a nestbox aiming
for another nestbox, consequently entering the corresponding transit box. It is constructed from the data as the inverse of
the mean leaving time per box.
 \item $\Theta$ is the \textit{transition rate}, a square matrix of length $B$. This is the rate at which an agent leaves a transit box to
enter in the corresponding destination nestbox. Is is constructed as the inverse of the mean transit time from any box to any
other box, including itself.
 \item $\Pi$ is the \textit{transition frequency}, a square matrix of length $B$. This decides of which transition actually occurs when
an agent is leaving a nestbox for another one, and accounts for the fact that transition probabilities between boxes are not equal.
$\Pi$ is row-stochastic, i.e. its row sum is always 1.
\end{itemize}

\subsubsection{Master equation}
In the mean field approximation, this $B$-body system can be considered as a $1$-body \textit{occupation density field} and thus
studied analytically by means of a system of master equations. For each state $k$, the probability of being in such a state
after a time $t$ depends on the transition rates to and from this state and the occupation density of all the other states,
in other words the variation of the density of agents in a particular state is the density of agents coming from other states
to which is subtracted the density of agents leaving for other states:
\begin{equation}
\displaystyle
 \forall k \in \{1,\dots,B\},\qquad \frac{\partial P_k}{\partial t} = \displaystyle\sum_{b=1}^B (T_{kb}P_k - T_{bk}P_b) \;  ,
\end{equation}

where $T$ is a $(B+B^2)$-square matrix of transition rates from any state to any other state. In our case, $T$ is a composite
of $\lambda$, $\Theta$ and $\Pi$, so that:

\begin{eqnarray}
\displaystyle
  \forall \{i,j\} \in \{1, \dots ,B+B^2\}^2, 	& 	& \forall n \in \{1,...,B\}, \nonumber \\
   T_{ij}					&   =   & \left\{
							    \begin{array}{ll}
							  \Pi_{i,j-B \cdot i} \cdot \lambda_i, & j \in \{B\cdot i+1,\ldots,B\cdot (i+1)\} \\
							  \Theta_{i,j}, & i = B\cdot n + j \\
							    0,         & \textnormal{otherwise}
							  \end{array} \right. \nonumber \\
						&	&
\end{eqnarray}



As there is only a restricted set of states reachable from any state (in fact, exactly $B$ from the states corresponding
to nestboxes and exactly $1$ from those corresponding to transit boxes), $T$ is mostly sparse.

\subsubsection{Analytical approach to the stationary distribution}
$T$ does not formally describe a Markov chain, since it is not row-stochastic. We have to construct a row-stochastic
transition matrix from $T$, in order to set a fixed timescale for our system: at each time step, a transition occurs.
For all $i$ and $j$ $\in \mathbb{N}^2$, $T_{i,j}$ is the joint probability for an agent of leaving box $i$ and entering
box $j$ thereafter. $\forall i$, $\sum_i T_{i,j}$ is the joint probability of leaving box $i$ and entering any box, that
is simply the probability of leaving box $i$. We need to include in the matrix $T$ the possibility that an agent does not
leave the box it is in after one timestep. If no leaving event has a rate (any of $T$ row-sums) higher than $1$s$^{-1}$,
there is a straightforward manner to convert the known transition rates to transition probabilities. We set our timestep
to $1$s and copy all elements of $T$ into a new matrix $S$. Since the rows of $S$ do not sum up to $1$, we define the
probability for an agent of staying in the same state as one minus the sum of all probabilities of leaving to any other
state, i.e. we obtain the stochastic matrix $S$ as follows:

\begin{equation}
\displaystyle
    S_{ij} = \left\{
      \begin{array}{ll}
	T_{ij},			& i \neq j \\
	1-\sum_k{T_{ik}},	& i = k \;  .
  \end{array} \right.
\end{equation}

We observe that in our case, since the mice stay a rather long time in a box, the leaving rate from each box is well below 1, so
the abovementioned condition applies. In the case where some events have a rate higher than $1$s$^{-1}$, we would have to
normalise all rates to the highest and follow the same process. In the end, if $i$ is the box with the fastest leaving rate,
$S_{i,i}=0$: the system's timescale has been adjusted to the fastest-happening event.

\begin{figure}
\centering
\includegraphics[width=11cm]{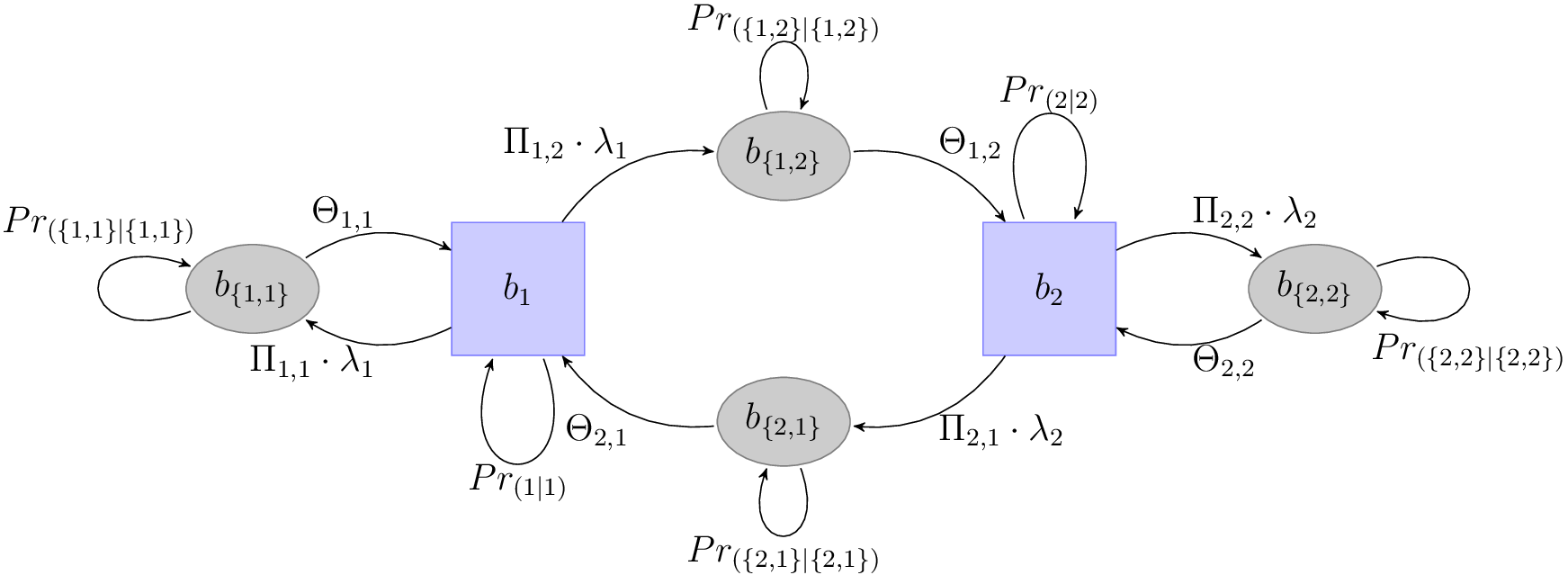}
\caption{State diagram of our modified Markov chain, so that for any two states $i$ and $j$, $S_{i,j}=Pr(j|i)$ is the transition
probability from $i$ to $j$. Note the additional transitions from any state $k$ back to the same state, $Pr_{(k|k)} = S_{k,k} =
1-\sum_j{\Pi_{i,j}} \cdot \lambda_i, \forall k\leq B$ and $Pr_{(k|k)} = S_{k,k} = 1-\Theta_{i,j}, k=B \cdot i + j, \forall k > B$.}
\label{fig:markovchain2}
\end{figure}


$S$ now describes a Markov chain that is slightly different from that of Fig.~\ref{fig:markovchain1}, as shown in
Fig.~\ref{fig:markovchain2}. The probability of moving from any state $i$ to any state $j$ during one time step is
$Pr(j|i) = S_{i,j}$. Let $d$ be the row vector of occupation density in all $B+B^2$ states. For any time $t$, we can write:

\begin{equation}
\displaystyle
 d(t+1) = d(t) \times S \;  ,
\label{eq:densityevol}
\end{equation}
where $d(t+1)$ is the occupation density vector after one arbitrary timestep. By recurrence, we can generalise
Eq.~\ref{eq:densityevol} to express the occupation density after $n$ timesteps:

\begin{equation}
\displaystyle
 d(t_n) = d_0 \times S^n \;  .
\end{equation}

We want to find the stationary probability vector $\delta$, i.e. the occupation density vector that does not change under
application of the transition matrix. As stated by the Perron-Frobenius theorem (\cite{seneta2006}), as our matrix is not
strictly positive, there are several corresponding vectors (the eigenvectors corresponding to the dominant eigenvalue).
We can however obtain a suitable $\delta$ as the limit of the application of $S$ many times on any non-zero initial
distribution. Let $d_0 \in \mathbb{R}^B$ characterise the initial distribution where all the occupation density is
concentrated in box $1$ and then slowly diffuses into all other boxes, $d_0 = \left[1,0,\dots,0\right]$. We can
numerically calculate $\delta$:
\begin{equation}
\displaystyle
 \delta = \lim_{k \to \infty} d_0 \times S^{k} \;  ,
\label{eq:deltalimit}
\end{equation}
Each of the first $B$ coefficients of $\delta$ is the stationary occupation density of a nestbox, normalised to 1.
To relate it to our observed occupation times, we would have to multiply each coefficient of the vector by the number
of timesteps (in our case the number of seconds) the simulation covers, and the mean number of agents in the system
(see~\ref{sec:sim-results}).

We compare the computed $\delta$ to our observed average occupation density after 1 hour (in which case the occupation densities
are given by $\delta_{3600} = d_0 \times S^{3600}$), 1 day, 1 week, 6 months and 1 year to observe the progressive establishment
of a stationary regime, in which all boxes have gotten a stable occupation density. Table~\ref{tab:occupations} shows the
corresponding Pearson's correlation coefficients and their associated p-values. We observe that the asymptotic regime is reached
in about 1 month.

\begin{table}
\centering
  \begin{tabular}{ | p{1.5cm} | c | c | }
    \hline
    \textit{t}	& \textit{$\rho$}	& \textit{p}	\\ \hline
    1 hour	& $-0.0428$		& $0.793$			\\ \hline
    1 day	& $0.1778$		& $0.272$			\\ \hline
    1 week	& $0.7248$		& $1.2 \cdot 10^{-7}$		\\ \hline
    1 month	& $0.9972$		& $2 \cdot 10^{-44}$		\\ \hline
    6 months	& $0.9995$		& $ < 10^{-50}$			\\ \hline
    1 year	& $0.9995$		& $ < 10^{-50}$			\\ \hline
  \end{tabular}
\caption{Comparison of the experimental average occupation density in the 40 nestboxes with the corresponding computed values
from an initial distribution where all the density is concentrated in box $1$, after a time $t$.}
\label{tab:occupations}
\end{table}

These results confirm that the Markov assumption holds well in our problem with regard to the average box occupation densities
over time. However, this analytical approach solely describes the ``diffusion'' of occupation densities, without characterising
the actual encounters that happen inside the nestboxes visited. For this reason, we use a multi-agent simulation to render the
social aspects of the system. Based on the results from the analytical approach, we can expect the simulation to yield (at least
for the box occupation frequencies) results similar to our experimental data.

\section{Stochastic simulation technique}
We use a stochastic simulation technique (\cite{schweitzer2007}) by defining the simulation time $\Delta t$ as a random variable,
depending on the current state of all agents: given N agents, each having their own transition time $t_i$ and transition rate
$f_i=1/t_i$,
we define the system's mean transition rate as $f_m = \frac{\sum_{i=1}^{N}f_i}{N}$, and transition time $t_m = 1/f_m$. This is the mean
frequency at which the system is expected to change. For each iteration, we sample the actual time step from an exponential
distribution with parameter (mean value) $t_m$, so that $\Delta t \sim \exp(t_m)$. We then sample the agent $k$ whose state is
changing (the Markov process that occurs) from a uniform distribution with regard to the transition frequencies, in
such a way:
\begin{equation}
\displaystyle
\forall p \sim U(0,1), \quad \exists k \in \{1,\dots,B\},\quad p \in \left[\sum_{i=1}^{k-1}P_i, \sum_{i=1}^{k}P_i\right[ \;  ,
\label{eq:lifetimesampling}
\end{equation}
where $P_i$ is the probability that process $i$ occurs ($\sum_{i=1}^{B}P_i = 1$). In other terms, $\Delta t$ indicates when
a change in the system is happening and $k$ indicates what this change is, according to each process's transition probability
matrix.

We define a constant number of agents in the system, equal to the average number of mice detected per day at the barn. We obtain
the leaving and transition rates as the inverse of the average leaving and transition times (extracted from the data) from
each box to each other box. For sampling reasons (see Eq.~\ref{eq:lifetimesampling}), we normalise those matrices so that their
sum is 1. We initially distribute the agents according to the occupation rate (in or out of a box) and the occupation preferencies.
The simulation timerange is set to the same period as that covered by the final dataset covers: Jan. 1, 2008 to Dec. 31, 2009.

\subsection{Results}
\label{sec:sim-results}
\begin{figure}
\centering
\includegraphics[width=5cm, angle=-90]{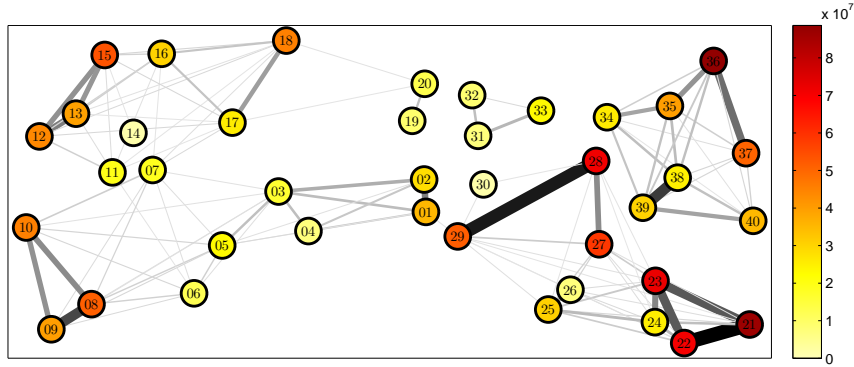}
\caption{Box occupation pattern and traffic between boxes, closely matching the pattern observed in Fig.~\ref{fig:boxoccupation}
(same legend applies).}
\label{fig:boxoccupationsim}
\end{figure}

We compare the occupation and movement pattern from the simulation (Fig.~\ref{fig:boxoccupationsim}) to that of the experimental
data (Fig.~\ref{fig:boxoccupation}). The simulation output resembles what we observed in the data: the
comparison between occupation rates and transit counts yield high Pearson's correlation coefficients, $\rho=0.975$ ($P=2.4 \cdot
10^{-26}$) and $\rho=0.993$ ($P<10^{-50}$), respectively.

\begin{figure}
\centering
\includegraphics[width=6cm]{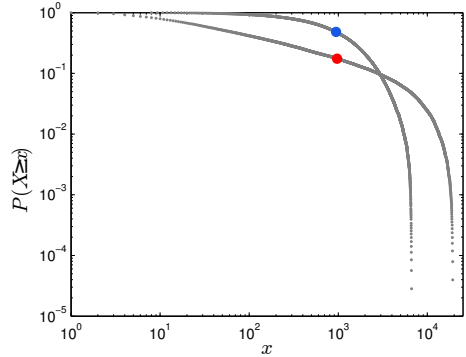}
\caption{Comparison of the cumulative probability distribution of leaving times from nestbox 1 in the real data (curve with the red dot)
and the simulation output (curve with the blue dot), with the leaving times given in seconds. Although the means of both distributions
(the coloured dots) are almost identical, their shapes are largely dissimilar (two-sample K-S test, $P<10^{-50}$).}
\label{fig:leavingtimes}
\end{figure}

However, when one looks at a more detailed description level, these similarities disappear: as shown in
Table~\ref{tab:metrics}, the model produces a slight overestimation of the agents' activity (this may come
from an overestimation of the number of agents in the barn by the daily mean) and more importantly a larger underestimation of the
agents' social behaviour (expressed here through the mean duration of a social contact, defined as an overlapping stay in the same
box with any other agent). Likewise, and although the mean leaving times from nestboxes are largely similar (Pearson's $\rho=0.999$
over the 40 nestboxes), their respective distributions differ: compared using a Kolmogorov-Smirnov test, the test yields a p-value
$< 0.05$ (continuous distributions significantly different) for 39 out of the 40 nestboxes. This is illustrated in
Fig.~\ref{fig:leavingtimes} for nestbox 1, which compares the real and simulated data. It is also interesting to note that although the
mean number of meetings occurring during one stay in a box is slightly less in the simulation output than in the data, the comparison
of the distribution of this ratio accross all boxes does not yield a significant difference (two-sample K-S test, $P=0.967$).

\begin{table}
\centering
  \begin{tabular}{   p{6cm}   c   c  }
    ~Metric					& ~Data    & ~Model
  \end{tabular}
  \begin{tabular}{ | p{6cm} | c | c |}
    \hline
    Total number of agents			& 76      & 508     \\ \hline
    Mean number of agents observed per day	& 76.05   & 76      \\ \hline
    Number of stays in a box per day*		& 21.88   & 35.44   \\ \hline
    Number of social contacts per day*		& 38.29   & 50.88   \\ \hline
    Duration of a stay (s)*			& 803.2   & 700.0   \\ \hline
    Duration of a social contact (s)*		& 1085.7  & 368.4   \\ \hline
    Mean number of meetings per stay		& 0.738	  & 0.683   \\ \hline
    Number of nestboxes visited per day*	& 2.61	  & 6.58    \\ \hline
  \end{tabular}
\\ \hspace{2cm} \raggedright{*average value per agent}
\caption{Comparison of some metrics between the experimental data and the simulation output}
\label{tab:metrics}
\end{table}

\subsection{Interpretation}
\subsubsection{Stochastic dimension of the system}
We observed in \ref{sec:sim-results} that some global properties of the system can be reproduced using our Markov chain model.
These include box occupation density, intensity of interbox transit, and propensity of the agents to move to
a box regardless of its current occupation (number and identity of other agents), as shown by the lack of significant difference
in the ratio meetings/stays between the real and simulated data.

The similarity in the occupation and transit patterns is only little surprising since the simulation's leaving and transition rates
were obtained from the data, but it validates our assumption on the Markovian nature of the system. Moreover, it indicates that the
collective dynamics of purely stochastic processes are sufficient to reproduce macro-level properties of a social system, i.e. that
no special cognitive ability is required in order to reproduce the observed spatial dynamics of box occupation and interbox transit.
Furthermore, from the similarity in the observed and simulated number of encounters per stay in a box it appears that, in order to
reproduce the meeting frequency of the agents it is sufficient to consider that an agent's motivation as to which box to visit next
depends solely on stochastic factors.

\subsubsection{Influence of social parameters}
However, a detailed insight into the simulation data reveals discrepancies with regard to the social dimension of the system:
it seems that although agents do not deliberately decide of their encounters with other agents, the output (expressed through
the duration) of those encounters depends largely on non-stochastic factors, such as presumably the common social past of the
agents and their current degree of social preference for each other. Likewise, the number of nestboxes visited daily differs
largely in the experimental and simulated data (two-sample K-S test, $P< 10^{-50}$). We argue that the discrepancies observed
come from so-called social agent-level interactions, mitigating the macro-level pattern observed earlier. This is to be corrected
by introducing in the model further assumptions derived from behavioural science.


\section{Discussion}
\label{sec:discussion}
We have shown in this paper that it is possible to reproduce some of the collective dynamics of an animal society by using stochastic
processes that make no call whatsoever to social behaviour. We used a top down approach and defined a very simple rule of movement
in the barn, governed by only one parameter (the time to next move). We chose not to fully describe the system by using in
our model design several simplifications, namely:
\begin{itemize}
\item we ignored all the individual characteristics of the agents (e.g., sex differences).
\item we assumed that the lifespan of an agent was infinite (although the lifespan of a wild house mouse averages well below 2 years,
see \cite{manser2009}).
\item we considered that an agent's behaviour was not affected by its age nor its past experience.
\item we finally included no consideration for the social interactions that can occur each time two mice encounter within or without
a box, although house mice are a cooperative breeding species and it has been long known that social aspects play a crucial role in
their behaviour (\cite{brown1953}, \cite{brain1989}).
\end{itemize}
Notwithstanding, we observed that the result of our stochastic simulation mat\-ches closely the macro-level properties of the real
dataset. This is clearly encouraging in the search for a more advanced cross-species model that could lead to a better and wider
understanding of animal sociality.

However, we also identified drawbacks in our approach, as illustrated in the fact that the distribution of leaving and transit times
we assume is not the same as the one we observe in reality, or that the average meeting time that comes out of our model is well
below the observed average meeting time. We also observe that the traffic per agent is concentrated among a smaller subset of boxes
than it is expected in our model, which is a strong hint to territoriality. If the laws of social behaviour in wild house mice were
as simple as we assumed here in a first step, they would lead to easily identifiable regular patterns, as it is often the case in
natural phenomena (\cite{murray2003}). But we obviously have to dig a bit deeper in the assumptions we make
on the social nature of the mice's behaviour. For a start, we could propose a refinement of our model where the likelihood of an agent
travelling to a certain box depends on whether this box belong to its territory or not. Furthermore, we could assume that the leaving
time of an agent once its has entered a box is a function of the number of the box's occupation density, i.e. there is a preferred
number of partners that fosters cooperation (for example: zero is not enough, but three is too much). Furthermore, the identity of
the agents encountered is certainly a factor of primary importance for the duration and the outcome of the meeting, as hinted at in
recent experimental studies (\cite{weidt2008}) on the same species. Therefore, introducing notions of rational choice theory with regard
to past social experience and expected fitness benefits in our model would likely be a great step forward in the reproduction of the
diversity of behavioural patterns we observe.

This contribution presents an application of a stochastic multi-agent model to the field of animal behaviour. It is considered a
first step towards more comprehensive research aiming to bring new insights to the field of behavioural sciences. Forthcoming
studies are certainly promising in what they can bring to the general understanding of how and why animals think, and behave,
socially.


\bibliographystyle{plain}
\bibliography{paper_eccs}

\end{document}